\documentclass[10pt,a4paper]{article}
\usepackage[font={small}]{caption}
\usepackage[margin=1in]{geometry}
\usepackage{xcolor}
\usepackage{changebar}
\usepackage{graphicx}

\def\opex{ Opt.\ Express }

\def\apl{ Appl.\ Phys.\ Lett.\ }

\title{Electro-optical switching by liquid-crystal controlled metasurfaces}

\author{Manuel Decker$^{1,*}$, Christian Kremers$^2$, Alexander Minovich$^{1}$, Isabelle Staude$^{1}$,\\ Andrey E. Miroshnichenko$^1$, Dmitry Chigrin$^2$, Dragomir N. Neshev$^1$,\\ Chennupati Jagadish$^3$, and Yuri S. Kivshar$^1$}

\begin{document}
\maketitle
\begin{center}
\noindent $^1$Nonlinear Physics Centre and Centre for Ultrahigh Bandwidth Devices for Optical Systems (CUDOS), \\
Research School of Physics and Engineering, The Australian National University, \\ Canberra ACT 0200, Australia\\
$^2$Institute of High-Frequency and Communication Technology, University of Wuppertal, Germany\\
$^3$Department of Electronic Materials Engineering, Research School of Physics and Engineering,\\ The Australian National University, Canberra ACT 0200, Australia\\

\noindent $^*$manuel.decker@anu.edu.au\\ 
\end{center}

\begin{abstract}
We study the optical response of a metamaterial surface created by a lattice of split-ring resonators covered with a nematic liquid crystal and demonstrate millisecond timescale switching between electric and magnetic resonances of the metasurface. This is achieved due to a high sensitivity of liquid-crystal molecular reorientation to the symmetry of the metasurface as well as to the presence of a bias electric field. Our experiments are complemented by numerical simulations of the liquid-crystal reorientation.
\end{abstract}


\section{Introduction}
\vspace{-2mm}
Many applications of photonic structures, such as photonic crystals and metamaterials, require the ability to dynamically change their properties on short timescales. For metamaterials, several tuning mechanisms utilizing phase-change materials~\cite{DriscollBasovScience2009,SamsonZheludevAPL2010}, semiconductors~\cite{ChenPadillaNatPhot2008} and liquid crystals (LCs)~\cite{ZhaoZhangAPL2007, XiaoShalaevAPL2009, KangWuOE2010, ZhangLippensOE2011, MinovichKivsharAPL2012} have been explored so far, however, the choice of LC infiltration is especially attractive because of the large optical anisotropy of the LC molecules. Utilizing this property, LCs allow for several different approaches of tunability, {\textit e.g.} by changing temperature, by applying an external electric or magnetic field, and even all-optical tuning by employing a strong nonlinear response of metamaterials (see review paper~\cite{KhooMaIEEE2010}).

To date, most works have focussed on the tunability of microwave metamaterials infiltrated with nematic LCs. For example, it was demonstrated that an external electric field changes the reorientation of liquid-crystal molecules and leads to an effective index change for an array of split-ring resonators (SRRs)~\cite{ZhaoZhangAPL2007} and microwave fishnet metamaterials~\cite{ZhangLippensOE2011}. The realization of LC tunability of metamaterials in the near infrared and optical regime is a much harder task, since the effects of molecular anchoring to the nanostructured surfaces becomes important when the dimensions of the individual meta-atoms become comparable to the size of the LC molecules -- this is particularly the case for optical metamaterials and has a significant impact on the tunability of optical metamaterials. Nevertheless, thermal~\cite{XiaoShalaevAPL2009} and UV-irradiation induced tunability~\cite{KangWuOE2010} of optical metamaterials has been experimentally tested, showing a slow temporal response. However, all-optical control of fishnet metamaterials infiltrated with LCs was studied experimentally only recently~\cite{MinovichKivsharAPL2012}.

In this paper, we demonstrate a novel feature of planar metamaterials strongly coupled to the LC molecules. By making use of the strong anchoring of the LC to the metallic elements of the surface, we imprint the metasurface symmetry onto the LC distribution. The concept of metasurface was introduced recently in the context of a two-dimensional lattice of plasmonic nanoantennas with spatially varying phase response and subwavelength separation~\cite{GenevetGaburroScience2011, NiShalaevScience2012}. Such metasurfaces allow for engineered light reflection due to a linear phase variation along the interface. In our case, however, the metasurface based on SRRs~\cite{LindenSoukoulisScience2004, DeckerWegenerPRB2011} can be either electric or magnetic, which defines the transmission properties of the LC-metamaterial cell. As such, we demonstrate dynamic switching between electric and magnetic resonances of the SRR metamaterial on a millisecond timescale and show that this effect is a result of the reorientation of the LC molecules in the presence of a bias electric field. Our scheme opens up new opportunities for fast switching between electric and magnetic metasurfaces, which also allows for dynamic control of light reflection.

\section{Characterization of the Liquid-Crystal/Metamaterial Cell}
\begin{figure}[hb]
  \centering
  \includegraphics[width=12.5cm]{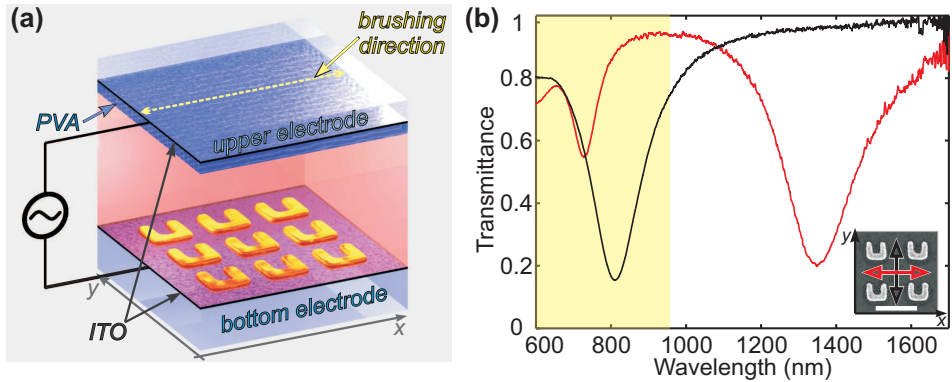}
  \vspace{-2mm}
  \caption{(a) Artistic view of the LC cell. The SRR metamaterial is processed on top of the bottom electrode while the upper electrode is covered with an alignment layer of mechanically brushed PVA. An AC power supply is connected to the conductive ITO films (plotted in dark gray). (b) Experimental transmittance spectra of the bare SRR metamaterial sample for {\textit x}- (red) and {\textit y}- (black) polarization of the incident light. The yellow area highlights the spectral range relevant for our experiment.}
  \vspace{-3mm}
  \label{setup}
\end{figure}

\vspace{-2mm}
The centerpiece of our LC-metamaterial cell shown in Fig.\,\ref{setup}(a) is an array of gold SRRs with a lattice spacing of $300\,{\rm nm}$ on a glass substrate covered with $5\,{\rm nm}$ of Indium-Tin-Oxide (ITO) fabricated by a standard electron-beam lithography (EBL) process. A close-up scanning-electron micrograph of our SRR metamaterial is shown in the inset of Fig.\,\ref{setup}(b). The lateral dimensions of the individual SRR meta-atoms are $l_{x}\approx 138\,{\rm nm}$, $l_{y}\approx 124\,{\rm nm}$ and the line width is $w\approx 45\,{\rm nm}$. The SRR thickness is $25\,{\rm nm}$.
Figure\,\ref{setup}(b) also shows the measured linear-optical transmittance spectra of the SRR metasurface without the LC. The (fundamental and higher-order) magnetic resonances of the SRR metamaterial are excited for {\textit x}-polarization of light ($\lambda_{m0}\approx 1350\,{\rm nm}$, $\lambda_{m1}\approx 765\,{\rm nm}$, and $\lambda_{m2}\approx 570\,{\rm nm}$) while the electric resonance ($\lambda_{e0}\approx 805\,{\rm nm}$) is excited for {\textit y}-polarization only [see red and black arrows in Fig.\,\ref{setup}(b)]. The transparent conductive ITO layer prevents charge accumulation during EBL, and also serves as one of the two electrodes needed to apply the electric potential to the LC cell. The second ITO-covered glass substrate serves as the top electrode of the LC cell. To allow for a defined pre-alignment of the LC molecules at this top electrode we additionally spin-coat a $200\,{\rm nm}$-thin layer of polyvinyl alcohol (PVA) on top of the ITO layer and mechanically brush the PVA to obtain a preferred direction for LC pre-alignment [see yellow arrow in Fig.\,\ref{setup}(a)]. No LC pre-alignment layer has been used on the metasurface.
Finally, the LC cell is assembled by placing a $31\,{\rm \mu m}$-thick spacer between the metamaterial substrate and the second PVA-coated electrode. The exact height of the spacing between the two electrodes is derived from measured Fabry-Perot oscillations of the LC cell without the LC. Importantly, the upper electrode is placed in a fashion that the brushing of the PVA layer and, hence, the pre-alignment of the LC molecules at the PVA surface is oriented perpendicular to the mirror plane of the SRRs on the bottom electrode [see Fig.\,\ref{setup}(a)]. After infiltration with the liquid crystal `Licristal E7' from Merck at $T=90^\circ C$, the LC cell is then mounted in a home-built white-light transmittance setup and connected to an adjustable function generator that provides a sinusoidal AC voltage with 1kHz frequency. In our experiments, the light from a halogen lamp first passes a polarizer to create linear polarization and is consecutively focussed onto the SRR metamaterial. Importantly, the incident light travels through the bulk LC before impinging onto the SRR metasurface. The SRRs in the metamaterial sample are oriented such that the magnetic resonances are excited by {\textit x}-polarization of light [indicated by the red arrow in the inset of Fig.\,\ref{setup}(b)]. After transmission through the sample, the emerging light spectrum is detected with an Ocean Optics spectrum analyzer. Alternatively, a CCD image of the SRR-metamaterial sample can be recorded. An additional polarizer can be inserted into the detection path to analyze the output polarization state of light.

\section{Switching of Metamaterial Resonances}
\begin{figure}[hb]
  \centering
  \includegraphics[width=12.5cm]{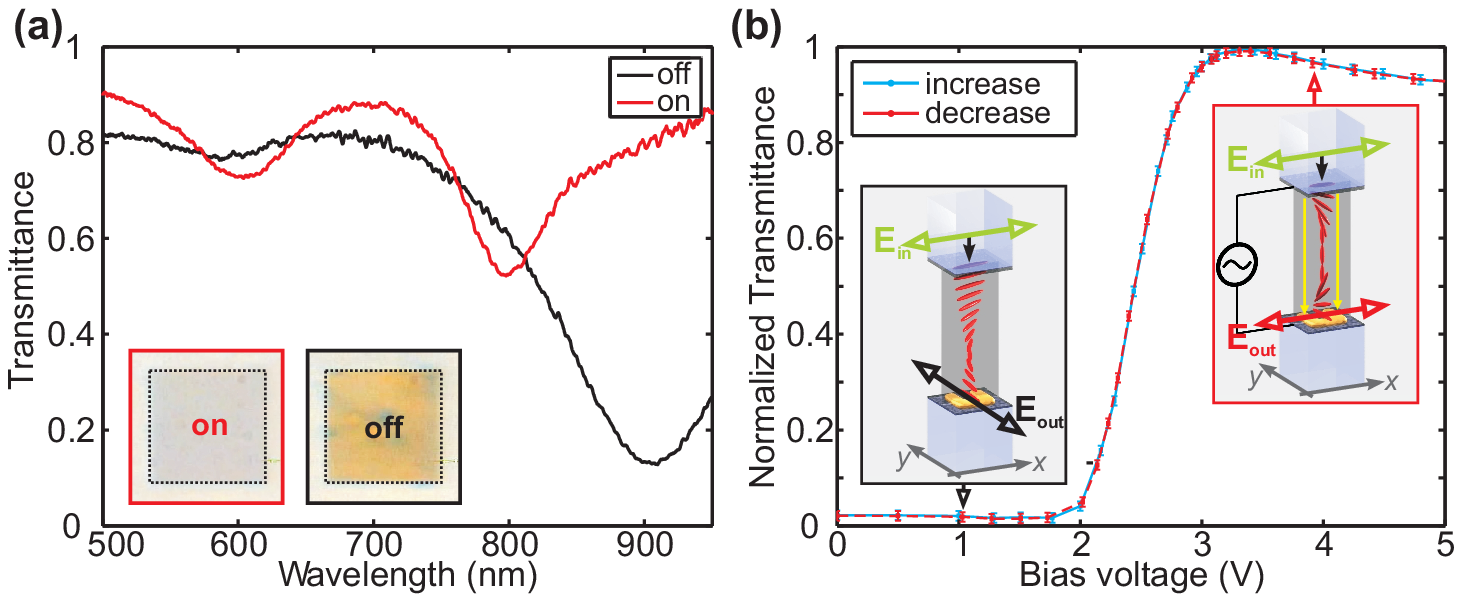}
  \vspace{-2mm}
  \caption{(a) Experimental transmittance spectra of the LC cell with no voltage applied (`OFF' state, black) and for $V=6\,{\rm V}$ (`ON' state, red). Insets: CCD images of the metamaterial area in the `OFF' and `ON' state.(b) Threshold behavior of the switching process for increasing (blue) and consecutive decreasing (red) voltages. The left inset depicts the (helical) LC distribution in the `OFF' state, the right inset shows the situation in the `ON' state (no helical distribution). The incident and output light polarizations are indicated as green and red/black arrows, respectively.}
  \vspace{-3mm}
  \label{experiment}
\end{figure}

\vspace{-2mm}
In our experiment we fix the incident polarization to {\textit x}-polarization and record the transmittance spectrum (without the second polarizer in the detection path) without a bias voltage ($V=0\,{\rm V}$, `OFF' state) and with a bias voltage of $V=6\,{\rm V}$ (`ON' state) at the two electrodes. The results are shown in Fig.\,\ref{experiment}\,(a). In this configuration the electric resonance at $\lambda_{e0}\approx 900\,{\rm nm}$ is excited in the `OFF' state. When changing from `OFF' to `ON' a clear change in the transmittance spectrum occurs and the magnetic resonances at $\lambda_{m1}\approx 800\,{\rm nm}$ and $\lambda_{m2}\approx 600\,{\rm nm}$ are observed. Inversely, when changing from `ON' to `OFF' the original spectrum is restored again. Due to the presence of the LC, the resonances are red shifted compared to the spectra shown in Fig.\,\ref{setup}(b).
Remarkably, the change in spectrum when going from `OFF' to 'ON happens on the millisecond timescale while going from `ON' to OFF' takes about 1 second to restore the original spectrum. The switching process can also easily be observed in the corresponding CCD images of the SRR metamaterial sample shown in the insets in Fig.\,\ref{experiment}(a). While in the `OFF' state the metamaterial area appears in a yellow color it nearly becomes transparent in the `ON' state.
Having a closer look at the experimental data, one realizes that for {\textit x}-polarization of the incident light (before passing through the LC) the electric resonance at $\lambda_{e0}=900\,{\rm nm}$ is excited in the `OFF' state. Therefore, the incident {\textit x}-polarized light must have undergone a polarization rotation of 90 degrees upon travel through the bulk LC volume which, in turn, can only be explained by assuming that the LC molecules on the SRR metasurface are oriented perpendicular to the pre-aligned LC molecules on the brushed PVA surface. Only then we would expect them to relax from {\textit x}- (upper electrode) to {\textit y}- (bottom electrode) orientation forming a helical distribution in between [see left inset of Fig.\,\ref{experiment}(b)]. This helical distribution of the LC molecules results in the well-known adiabatic rotation of the incident polarization from {\textit x}- to {\textit y}-polarization. As a result, for {\textit x}-polarized incident light the electric resonance is excited in the `OFF' state and the emerging polarization state of light is then {\textit y}-polarized. The equivalent reasoning holds true for {\textit y}-polarized incident light and was also confirmed experimentally (not shown). Furthermore, when introducing the second polarizer in the detection path we verified that the output polarization of light is indeed rotated by 90 degrees.

When switching on the bias voltage, the LC molecules between the two electrodes are forced to reorient to be parallel to the electric field in the LC volume. As a consequence the helical distribution of the LC molecules is destroyed and the polarization state of light is conserved when travelling through the bulk LC [see right inset of Fig.\,\ref{experiment}(b)]. This results in the excitation of the (higher-order) magnetic resonances at $\lambda_{m1}=600\,{\rm nm}$ and $\lambda_{m2}=800\,{\rm nm}$ [red solid line in Fig.\,\ref{experiment}\,(a)]. Hence, we are able to dynamically and reproducably switch between electric and magnetic SRR-metamaterial resonances by applying a bias electric field to the electrodes of the LC cell. Since the reorientation of the LC molecules is driven by an external stimulus, the on-switching process takes place within 1 millisecond while the off-switching process (without an external stimulus) is purely governed by relaxation processes and inherently slower (about 1 second in this case). However, switching times can significantly be reduced by, \textit{e.g.}, reducing the thickness of the spacer layer, optimization of the LC mixture, and specifically designing the upper electrode to allow for an electrically driven off-switching process as well.

Finally, we studied the threshold behavior of the switching process and recorded the voltage dependent transmittance for {\textit y}-polarized incident light. In this configuration, the emerging light is {\textit x}-polarized in the `OFF' state, and no transmittance is observed in {\textit y}-polarization, while in the `ON' state the full spectrum is detected. The voltage dependent transmittance is then averaged over the whole spectral range. Clearly, the switching process for increasing (blue line) and decreasing (red dashed line) values of the applied voltage reproducably takes place in the voltage range of $2-3\,{\rm V}$ showing no hysteresis [see Fig.\ref{experiment}(b)]. Above $V=5\,{\rm V}$ the transmittance saturates and the spectra remain unchanged. Hence, for this case the LC molecules are completely reoriented by the electric field. Importantly, the overall behavior described above is not observed for the area adjacent to the SRR metamaterial so that any pre-alignment effects of the LC by the surface of the bare electrode (without metamaterial) can be excluded.

\section{Numerical Results}
\vspace{-2mm}
In order to support the interpretation of our experimental findings, particularly the behavior of the LC molecules, we simulate the reorientation process of the LC. The LC-director distribution, which minimizes the Frank-Oseen elastic energy, is obtained by solving the dynamic Euler-Lagrange equation numerically using an explicit two-step second-order accurate MacCormack scheme~\cite{KremersChigrin2012}. The LC-director distribution is obtained as self-consistent solution of the Euler-Lagrange equation and the generalised Poisson equation taking strong anchoring at the boundaries and the applied voltage into account~\cite{KremersChigrin2012}. The coupling between the two equations is implemented via the dielectric tensor field based on the LC-director distribution. Having calculated the equilibrium LC-director distribution the resulting dielectric tensor field is used in a finite integration time-domain Maxwell solver to calculate the optical response.
All simulations have been performed using the same equidistant mesh with a spacing of $\Delta x=5\,{\rm nm}$ and periodic boundary conditions in $x-y$ direction. The dimensions of an individual SRR used in the calculations are $l_x=140\,{\rm nm}$, $l_y=120\,{\rm nm}$, $w=45\,{\rm nm}$ and a SRR height of $25\,{\rm nm}$. To reduce the computational burden on the one hand and achieving a polarization rotation in the `OFF' state on the other hand (Mauguin condition for the twisted nematic cell~\cite{Chen2011}) we use a LC-cell thickness of $3\,{\rm\mu m}$ as a compromise. The LC cell is sandwiched between two $10\,{\rm nm}$-thick ITO layers ($\epsilon_{ITO}=3.2$). The computational domain is terminated above and below by a perfectly matched layer of $10\times\Delta x$. The direction of the LC molecules on the top ITO layer and the SRR metasurface is fixed to $x$- and $y$-orientation, respectively. In order to speed up the relaxation process, a voltage of $25\,{\rm V}$ has been applied between the two electrodes. The bulk elastic distortions of the liquid crystal are described by the Frank elastic constants $K_1 = 11.1 \,{\rm pN}$, $K_3 = 17.1 \,{\rm pN}$ (data sheet Merck Licristal E7), and $K_2 =  10.32 \,{\rm pN}$~\cite{PolakZumer1994}.The static and optical permittivities are set to $\epsilon_o^{stat} = 5.2$, $\epsilon_e^{stat}  = 19.3$, and $\epsilon_o^{dyn} = 2.3155$ and $\epsilon_e^{dyn} = 3.0527$, respectively (data sheet Merck Licristal E7). For the dielectric function of gold we use a Drude-Lorentz model with  
$\epsilon(\omega)=\epsilon_{\infty}-{\omega_p^2}/{\left(\omega^2+i\gamma_c\omega\right)}+\sum^{2}_{j=1}{\Delta\epsilon_j\omega_j^2}/{\left(\omega_j^2-\omega^2-i\gamma_j\omega\right)}$.
\begin{figure}[htb]
  \centering
  \includegraphics[width=12.5cm]{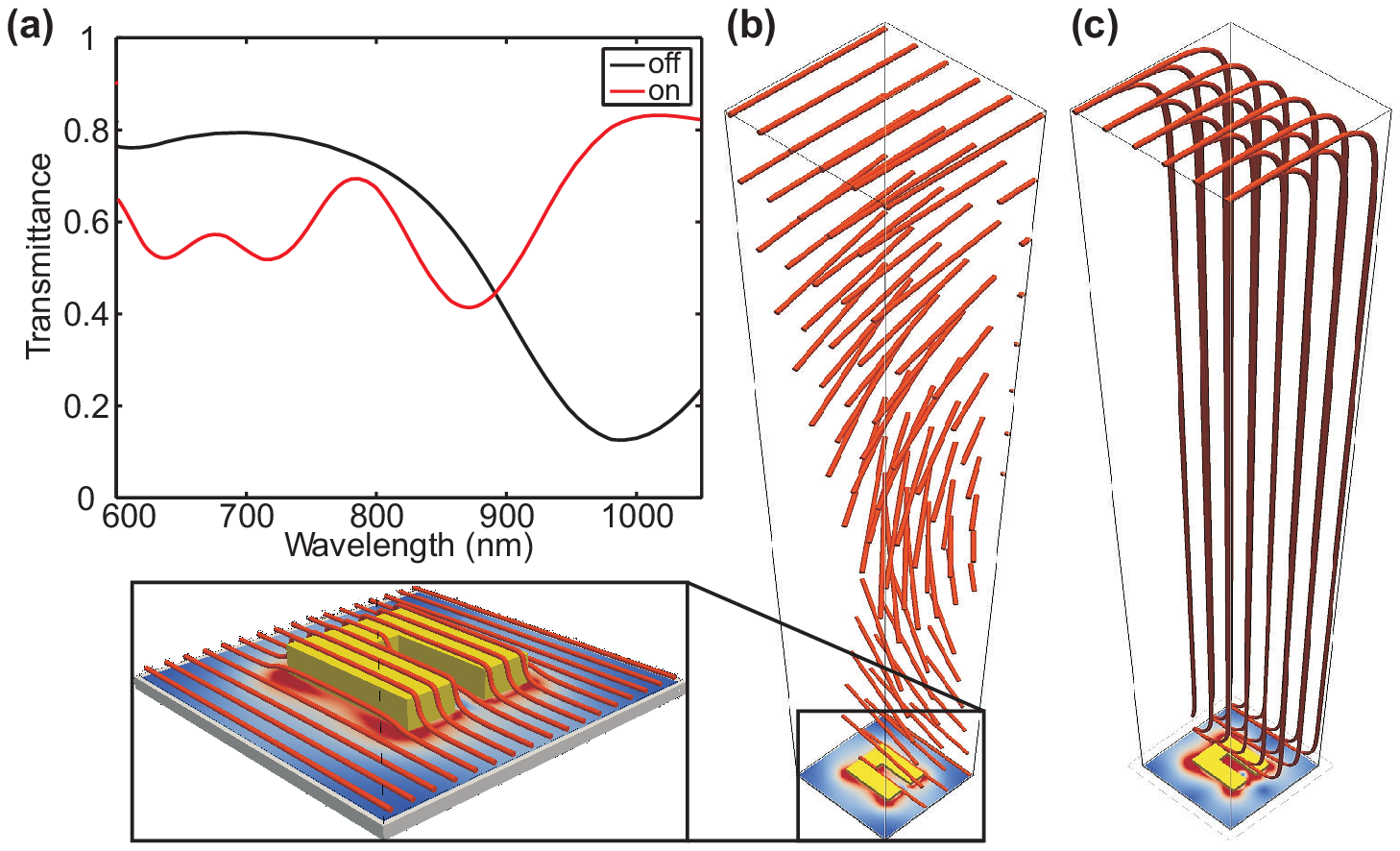}
  \vspace{-2mm}
  \caption{(a) Calculated transmittance spectra for the `OFF' and `ON' state in black and red, respectively. (b,c) Distribution of the LC directors in a $3 \mu m$-thick LC cell in the `OFF' and `ON' states, respectively. The orientation of the LC molecules is visualized as red streamline. The inset in the lower left shows a close up of the distibution of the LC molecules near and on the SRR metasurface. }
  \vspace{-3mm}
  \label{theory}
\end{figure}

The Drude parameters are $\epsilon_{\infty}=10.07$,$\omega_p=2\pi\times 2155 \,{\rm THz}$, $\gamma_c=2\pi\times 23 \,{\rm THz}$ and the two Lorentz functions are given by $\Delta\epsilon_1=0.25$,$\omega_1=2\pi\times 600 \,{\rm THz}$, $\gamma_1=2\pi\times 45 \,{\rm THz}$, and $\Delta\epsilon_2=0.15$,$\omega_2=2\pi\times 540 \,{\rm THz}$, $\gamma_2=2\pi\times 70 \,{\rm THz}$.
For better comparison of the experimental and numerical results, the Fabry-Perot fringes due to the reduced LC-cell thickness in the numerical spectra have been filtered out using Fourier smoothing. An excellent agreement between the simulated transmittance spectra in Fig.\,\ref{theory}(a) and the experimental spectra in Fig.\,\ref{experiment}(b) is found. The simulations also support our interpretation that in the `OFF' state the LC molecules form a helical arrangement which rotates the incident polarization of light by 90 degrees [see Fig.\,\ref{theory}(b)]. They also confirm the predicted reorientation of the LC molecules when applying a bias electric voltage above a characteristic threshold value [Fig.\,\ref{theory}(c)]. Due to a strong surface anchoring, reorientation of the LC molecules on the SRR metasurface is suppressed [see inset in Fig.\,\ref{theory}].

\section{Conclusions}
\vspace{-2mm}
We have demonstrated dynamic mode-switching between electric and magnetic resonances of a SRR-metasurface at near-infrared wavelengths using a nematic liquid crystal. We have attributed this effect to a high sensitivity of the LC molecules to the symmetry of the SRR metasurface that causes the LC to pre-align parallel to the mirror plane of the SRR metamaterial. Our findings have important implications for liquid-crystal tuning of plasmonic resonances since those experiments crucially rely on the ability to reorient the LC molecules within the near-fields of the plasmonic structure in the near-infrared spectral range. Hence, a profound understanding of the LC properties is indispensible for realizing those experiments.

After completing this manuscript we became aware of a recent paper~\cite{BuchnevFedotov2013} where a similar approach was applied to control a completely different type of plasmonic metamaterial.

\section{Acknowledgments}
\vspace{-2mm}
We thank I. C. Khoo for useful discussions and acknowledge the financial support from the Australian Research Council. The Fabrication Facilities used in this work have been supported by the ACT node of the Australian National Fabrication Facility (ANFF).

\end{document}